\documentclass[a4paper]{article}

\usepackage{type1cm}   
\usepackage{dsfont}         
\usepackage{makeidx}         % allows index generation
\usepackage{graphicx}        % standard LaTeX graphics tool
\usepackage{multicol}        % used for the two-column index
\usepackage[bottom]{footmisc}% places footnotes at page bottom

\usepackage{newtxtext}       % 
\usepackage{newtxmath}       % selects Times Roman as basic font
\usepackage{pict2e}

\usepackage{algorithm}% http://ctan.org/pkg/algorithms
\usepackage{algpseudocode}
\algtext*{EndIf}
\algtext*{EndFor}
\algtext*{EndWhile}
\usepackage{amsmath}
\usepackage{subcaption}
\usepackage{mathrsfs}
\usepackage{tikz}
\usepackage{mathbbol}

\newcommand{\wabs}[1]{{\left| {#1} \right|}}

\newcommand{\wcal}[1]{{\cal {#1}}}

\newcommand{\wds}[1]{{\mathds #1}}

\newcommand{\wfc}[2]{{#1}\!\left(#2\right)}

\newcommand{\whs}[1]{\hspace{#1cm}}

\newcommand{\wlr}[1]{\left({#1}\right)}
\newcommand{\wnorm}[1]{\left\|{#1}\right\|}
\newcommand{\wo}[1]{\overline{#1}}

\newcommand{\wref}[1]{$\wlr{\ref{#1}}$}

\newcommand{\wrn}[1]{{\mathds R}^{#1}}
\newcommand{\wrm}[1]{\mathrm{#1}}

\newcommand{\ws}[1]{\wcal{#1}}
\newcommand{\wset}[1]{\left\{ {#1} \right\}}
\newcommand{\wtr}[1]{\mathrm{T}}
\newcommand{\wu}[1]{\underline{#1}}
\newcommand{\wv}[1]{\mathbf{#1}}

\usepackage{authblk}
\title{Solving systems of inequalities in two variables with floating point arithmetic}
\author[1]{Walter F. Mascarenhas}
\affil[1]{Departamento de Computação, IME\\ Universidade de São Paulo, Brazil}
\date{\vspace{-5ex}}
\begin{document}

\maketitle

\begin{abstract}
From a theoretical point of view, finding the solution set of a
system of inequalities in only two variables is easy. However, if we want to 
get rigorous bounds on this set with floating point arithmetic,
in all possible cases, then things are not so simple due to
rounding errors. In this article we describe in detail
an efficient data structure to represent this solution set and 
an efficient and robust algorithm to build it
using floating point arithmetic. The data structure and the algorithm 
were developed  as a building block for the rigorous solution of
relevant practical problems. They were implemented  in \texttt{C++}
and the code was carefully tested. This code is available 
as supplementary material to the arxiv version of this article,
and it is distributed under the Mozilla Public License 2.0.
\end{abstract}

\section{Introduction}
\label{sec_intro}
We consider the representation and computation of the set $\ws{F} \subset \wrn{2}$ of points 
which satisfy the system of inequalities
\begin{equation}
\label{lp}
\begin{array}{ll}
a_i x + b_i y \geq c_i & \wrm{for} \ i = 1,\dots,n, \\
    x,y \geq 0. &
\end{array}
\end{equation}
This problem has applications in economics and computational geometry
\cite{CG}, but we developed the data structure and 
algorithm presented here for finding rigorous bounds
on the solutions of two dimensional nonlinear programming problems
using interval arithmetic \cite{Moore}.
For instance, when using Newton's method
to solve a nonlinear system of equations
\[
\wfc{f}{x} = 0  \whs{1} \wrm{for} \whs{1} f: \wrn{2} \rightarrow \wrn{2}
\]
with branch and bound, in each branch we have a
candidate set of solutions $\ws{S}$ described by a 
family of inequalities as in Equation \wref{lp}. We then
compute an interval matrix $\wv{A}$ containing the jacobian
matrix of $f$ for all $x \in \ws{S}$ and execute the 
interval arithmetic version of the Newton step
$\ws{S} \gets \ws{S}\cap \wlr{x_c - \wv{A}^{-1} \wfc{f}{x_c}}$
for some $x_c$ near to the center of $\ws{S}$. 
The Newton step  yields linear inequalities, which we use
to refine $\ws{S}$. For this refinement to be bullet proof 
we need data structures and algorithms like the ones presented here. 
Many relevant problems can be solved
this way, producing rigorous bounds on the solutions.
The algorithm can be used as a building block for finding the complex roots of polynomials,
the periodic orbits of chaotic systems \cite{Alex}
or rigorous bounds on the solutions of ordinary differential equations.
   
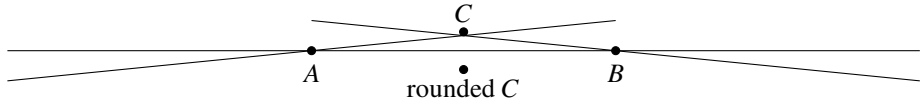
\begin{figure}[!h]
\begin{tikzpicture}[scale = 1]

\draw (0,0)--(12,0);
\draw (0,-0.4)--(8,0.4);
\draw (12,-0.4)--(4,0.4);

\filldraw[black] (4,0) circle(1.5pt);
\filldraw[black] (6,0.25) circle(1.5pt);
\filldraw[black] (8,0) circle(1.5pt);
\draw node at (6,0.5){$C$};
\draw node at (4,-0.3){$A$};
\draw node at (8,-0.3){$B$};
\draw node at (6,-0.5){rounded $C$};
\filldraw[black] (6,-0.25) circle(1.5pt);

\end{tikzpicture}
\caption{Rounding errors may lead to the conclusion that the
feasible region $ABC$ does not contain the vertex $C$, because
its rounded version lies below the line $AB$.}
\label{fig1}
\end{figure}

There are several algorithms for obtaining a reasonable 
representation of the feasible set $\ws{F}$ for the  
system of inequalities \wref{lp} 
using $\wfc{O}{n \log n}$  arithmetic operations \cite{CG}, 
but implementing them with floating
point arithmetic is not trivial due to
rounding errors. These errors lead to the worst kind of bug: 
the ones that occur only in rare situations and are difficult
to spot by testing. A simple example of 
what can go wrong is presented in Figure \ref{fig1}.
In fact, after much experience developing real world software for
computing Voronoi diagrams and Delaunay triangulations \cite{CG}, we
were convinced that it would be best to use exact arithmetic
instead of floating point arithmetic for this kind of problem, 
even knowing quite well that exact arithmetic is much more
expensive that floating point arithmetic.
The bugs caused by floating point arithmetic were overwhelming. 
In particular, the naive idea of using tolerances ($\epsilon$'s) 
does not work: in our experience, it is impossible to find the proper $\epsilon$'s
in a consistent way, which works in general.
Only recently we
came to the conclusion that it is possible to perform these
tasks with floating point arithmetic, provided that we 
use the techniques presented here and in the companion
article \cite{ES}.

For the kind of problems that we have in mind, it is 
acceptable to overestimate $\ws{F}$ a bit,
but we must not underestimate it. For instance,
if our algorithm indicates that $\ws{F} = \emptyset$
then it should be empty. On the other hand, it
is acceptable to reduce the right hand side of the constrains a bit. 
Therefore, in the next sections we describe an algorithm
with the following characteristics

\begin{itemize}
\item We assume that $\ws{F}$ is bounded, but it
may be a point, a segment or empty.  These cases 
cover all applications that we have in mind.
\item The algorithm finds a sharp approximation $\hat{\ws{F}}$ of $\ws{F}$, 
in the sense that $\hat{\ws{F}}$ is the exact feasible region for a slightly perturbed 
problem with constraints
$\tilde{a}_i, \tilde{b}_i$ and $\tilde{c}_i$ such that
\[
a_i x + b_i y \geq c_i \Rightarrow
\tilde{a}_i x + \tilde{b}_i y \geq \tilde{c}_i.
\]
In other words, $\ws{F} \subset \hat{\ws{F}}$ and the area of the set
$\hat{\ws{F}} \setminus \ws{F}$ is small.

\item When using a floating point type \texttt{T}, 
the data structure requires at most $\mu \ \wfc{\texttt{sizeof}}{\texttt{T}}$ 
bytes of memory, for a mild constant $\mu$.The algorithm
performs at most $\kappa n \wfc{\log}{n}$ sum, subtractions and 
multiplications, with a mild constant $\kappa$,
and uses at most $2n$ divisions.
\end{itemize}

In the next sections we describe the algorithm and the data structure
used to implement it. Section \ref{sec_build} gives a bird's eye view of 
the algorithm and points to the issues we face when translating this view into real code.
Section \ref{sec_repr} describes the data structure used to implement
the algorithm. Finally,
Section \ref{sec_test} emphasizes that we must be meticulous 
when testing the algorithm. Of course, an article like this
is no replacement for the real code when one wants to fully
understand the details. Such a code is available 
as supplementary material for the arxiv version of this article,
and is distributed under the Mozilla Public License 2.0. 
For practical reasons, the coded algorithm deviates a bit
from what we describe here, but ideas are the same.

\section{Computing $\ws{F}$}
\label{sec_build}

\begin{center}
\begin{figure}
\begin{tikzpicture}[scale = 1]

\filldraw[black] (0,0) circle(0.01pt);
\filldraw[black] (12,0) circle(0.01pt);
\filldraw[black] (0,4) circle(0.01pt);
\filldraw[black] (12,4) circle(0.01pt);

\draw (2,1)--(4,1)--(5,2)--(5,2.5)--(4.6,3.4)--(4,4)--(2,4)--(1,3)--(1,2.5)--(1.4,1.6)--(2,1);

\draw node at (2, 1.3){$B$};
\draw node at (4, 0.7){$C$};
\filldraw[black] (4.53,1.53) circle(1.5pt);
\draw node at (4.9, 1.53){$X$};

\draw node at (5.3,   2){$D$};
\draw node at (4.7, 2.5){$E$};
\draw node at (4.3, 3.2){$F$};
\draw node at (4,   3.7){$G$};
\draw node at (1.8,   4.3){$H$};
\filldraw[black] (2.5, 4) circle(1.5pt);
\draw node at (2.6, 4.3){$Y$};

\draw node at (1.3,   3){$I$};
\draw node at (1.3, 2.5){$J$};
\draw node at (1.25,1.45){$A$};

\filldraw[black] (2,     1) circle(1.5pt);
\filldraw[black] (4,     1) circle(1.5pt);
\filldraw[black] (5,     2) circle(1.5pt);
\filldraw[black] (5,   2.5) circle(1.5pt);
\filldraw[black] (4.6, 3.4) circle(1.5pt);
\filldraw[black] (4,     4) circle(1.5pt);
\filldraw[black] (2,     4) circle(1.5pt);
\filldraw[black] (1,     3) circle(1.5pt);
\filldraw[black] (1,   2.5) circle(1.5pt);
\filldraw[black] (1.4, 1.6) circle(1.5pt);

% A
\draw (0.2, 3.1)--(2.7,0); 
\draw node at (1.4 + 0.16 * 3.1 + 0.2, 1.6 + 0.16 * 2.5 + 0.1){$N$};
\draw[->](1.4 ,1.6)--(1.4 + 0.16 * 3.1, 1.6 + 0.16 * 2.5);

% const
\draw (2.5 - 0.1 * 3.3, 4 + 0.1 * 4)--(5.8, 0.0);
\filldraw[black] (2.5 + 0.9 * 3.3, 4 - 0.9 * 4) circle(1.5pt);
\draw[->](2.5 + 0.9 * 3.3, 4 - 0.9 * 4)--(2.5 + 0.9 * 3.3 + 0.16 * 3.1, 4 - 0.9 * 4 + 0.16 * 2.5);
\draw node at (2.5 + 0.9 * 3.3 + 0.16 * 3.1 + 0.2, 4 - 0.9 * 4 + 0.16 * 2.5 + 0.2){$N$};
\draw node at (2.5 + 0.9 * 3.3 - 0.2, 4 - 0.9 * 4 - 0.2){$K$};

% F
\draw (3.6, 4.6)--(5.3,  2.6);
\draw[->](4.6,3.4)--(4.6 + 0.16 * 3.1, 3.4 + 0.16 * 2.5);
\draw node at (4.6 + 0.16 * 3.1 + 0.2, 3.4 + 0.16 * 2.5 + 0.2){$N$};

\draw node at (9.5,4.2){normal vectors};

\draw node at (9.5,3.75){$BC$};
\draw[->] (9.5,2)--(9.5,3.5);

\draw node at (9.5,0.3){$GH$};
\draw[->] (9.5, 2)--(9.5,0.5);

\draw node at (11.2,2){$IJ$};
\draw[->] (9.5, 2)--(11,2);

\draw node at (7.6,2){$DE$};
\draw[->] (9.5,2)--(8,2);

\draw node at (8.2,3.3){$CD$};
\draw[->] (9.5, 2)--( 8.4, 3.1);

\draw node at (10.9,0.9){$HI$};
\draw[->] (9.5, 2)--(10.6, 0.9);

%(1,2.5)--(1.4,1.6) -> 0.4, -0.9 => 0.9, 0.4
%(1.4,1.6)--(2,1)   -> 0.6, -0.4 => 0.4  0.6

\draw node at (9.5 + 1.5 * 0.9 + 0.3, 2 + 1.5 * 0.4){$JA$};
\draw[->] (9.5,2)--(9.5 + 1.5 * 0.9, 2 + 1.5 * 0.4);

\draw node at (9.5 + 2.0 * 0.4 + 0.3, 2 + 2.0 * 0.6+ 0.2){$AB$};
\draw[->] (9.5,2)--(9.5 + 2.0 * 0.4, 2 + 2.0 * 0.6);

\draw node at (9.5 - 1.5 * 0.9 - 0.4, 2 - 1.5 * 0.4){$EF$};
\draw[->] (9.5,2)--(9.5 - 1.5 * 0.9, 2 - 1.5 * 0.4);

\draw node at (9.5 - 2.0 * 0.4 - 0.2, 2 - 2.0 * 0.6 - 0.2){$FG$};
\draw[->] (9.5,2)--(9.5 - 2.0 * 0.4, 2 - 2.0 * 0.6);

\draw node at (9.5 + 0.36 * 3.1 + 0.2, 2 + 0.36 * 2.5 + 0.1){$N$};
\draw[->](9.5,2)--(9.5 + 0.36 * 3.1, 2 + 0.36 * 2.5);

\draw node at (9.5 - 0.36 * 3.1 - 0.3, 2 - 0.36 * 2.5 - 0.1){$-N$};
\draw[->, dashed] (9.5,2)--(9.5 - 0.36 * 3.1, 2 - 0.36 * 2.5);

\end{tikzpicture}
\caption{Intersecting the feasible region with the half plane to the right
of the line passing through the point $K$. The vector $N$ is normal to
this line and points to the right.}
\label{figAlgo}
\end{figure}
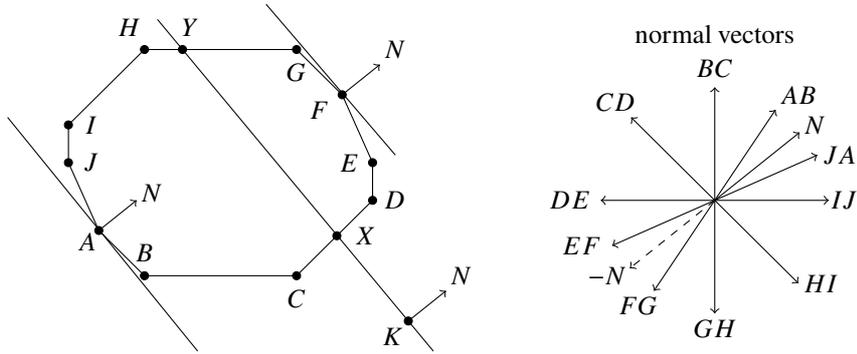
\end{center}

This section starts with a naive description of the
algorithm to compute the feasible region $\ws{F}$ 
for the system of inequalities \wref{lp}.
We then explain why this description is naive. We hope
that this presentation will motivate the data structure
we use to represent $\ws{F}$ in \texttt{C++} described
in the next section. 
In the problems with which we are concerned, 
$\ws{F}$ is contained in a box given by constraints 
\begin{equation}
\label{bound}
0 \leq x \leq m_x \whs{0.5} \wrm{and} \whs{0.5} 
0 \leq x \leq m_y 
\whs{0.5} \wrm{with} \whs{0.5}
m_x + m_y < \omega,
\end{equation}
where $\omega$ is the largest finite floating point value.
The algorithm assumes that Equation \wref{bound} holds and starts the construction 
of $\ws{F}$ from this box, and adds one constraint a time,
at the cost of $\wfc{O}{\log n}$ floating point operations per constraint.
From the geometric perspective of Figure \ref{figAlgo},
 adding a constraint $a x + b y \geq c$,
with  $a \neq 0$ say, corresponds to intersecting the
current feasible region with the half plane defined by the
line with normal vector $N := \wlr{a,b}$ containing
$K := \wlr{(c - b)/a, 1}$, and we proceed as follows:

\begin{enumerate}
\item[(i)] We keep a data structure with the inward normal vectors of the edges
of the feasible region sorted in counter clockwise order as in the right
of Figure \ref{figAlgo}. This data structure can be simply a vector
if we expect that the number $n$ of constraints will not be large, or 
a tree or other container with $\wfc{O}{\log n}$ cost for the
basic operations if we expect $n$ to be large. 
We perform two binary searches: one to locate
$N$ and another to locate $-N$. By doing so we find the 
vertex $A$ which minimizes $\wfc{f}{x,y} = a x + b y$ in $\ws{F}$ and
the vertex $F$ which maximizes $f$ in $\ws{F}$ (these are the vertices
of $\ws{F}$ which satisfy the Karush/Kuhn Tucker conditions for minimizing and
maximizing $f$ over $\ws{F}$.)
\item[(ii)] The vertices are also sorted in counter clockwise order, and
$f$ increases between $A$ and $F$ and decreases between $F$ and $A$.
We perform one binary search in $A...F$ to find the intersection point $X$
and another in $F...A$ to find $Y$,
again at a cost of $\wfc{O}{\log n}$ floating point operations.
\item[(iii)] We then have the new feasible region $XDEFGY$.
\end{enumerate}

This description of the algorithm is naive because it ignores
many details and corner cases. A proper description would prescribe, for example,
what to do if $X = Y = A$ or $X= Y = F$, or even $X = A$ and $Y = B$. 
In fact, taking proper care of these particular cases is what consumes
most of the time when coding this kind of algorithm. The naive case
described above is easy (see Section \ref{sec_test} for 
more challenging examples.) 
Moreover, as in some books about computational geometry
\cite{OR}, the naive description also assumes
that we can perform the three operations
below exactly, but doing so requires care when
using floating point arithmetic:

\begin{itemize}

%%%%%%%%%%%%%%%%%%%%%%%%%%%%%%%%%%%%%%%%%%%%%%%%
%%% normalizing normals
%%%%%%%%%%%%%%%%%%%%%%%%%%%%%%%%%%%%%%%%%%%%%%%%

\item[(i)] When finding the location of the normal $N$ in the set of normals,
we must decide whether $N$ comes before or after the normal $n$ to an edge, 
or whether it has the same direction as $n$ in degenerate cases. 
In algebraic terms, this reduces to the computation of the sign of the determinant 
\begin{equation}
\label{ang_order}
n_x N_y - n_y N_x,
\end{equation}
and the first hurdle we face is the possibility of overflow
in the products $n_x N_y$ and $n_y N_x$. We believe that the easiest way to 
handle this possibility is to normalize normals up front. 
When $a_i > \wabs{b_i}$, the idea is to replace the
constraint
\begin{equation}
\label{ci}
a_i x + b_i y \geq c_i
\end{equation}
by
\[
x + \frac{b_i}{a_i} y \geq \frac{c_i}{a_i}.
\]
Unfortunately, this cannot always be done exactly in floating point arithmetic,
and we need to round numbers consistently. It turns out that it is
simpler to do this consistent rounding when the rounding mode is upwards,
and our algorithm ensures that this is the rounding mode at its very beginning.
Recalling that $y \geq 0$, we can then relax the constraint \wref{ci} by 
replacing it by 
\begin{equation}
\label{cin}
x + \wlr{b_i \oslash a_i} y \geq - \wlr{ \wlr{-c_i} \oslash a_i },
\end{equation}
where by $x \oslash y$ we mean the floating point division of $x$ by $y$ rounding up.
The double negation in the right hand side of the modified constraint \wref{cin} 
is equivalent to rounding down $c_i/a_i$. 
As a result, by replacing the constraint \wref{ci} by the
constraint \wref{cin} we can add points to the feasible region, but no
points will be removed from it. The cases in which the assumption $a_i > \wabs{b_i}$
is not satisfied are handled analogously. Except for the trap $a_i = b_i = 0$, there
are eight possibilities, which correspond to the decomposition of the
interval $[0,2\pi)$ in eight disjoint semi open intervals of width $\pi /4$:
\[
a_i > b_i \geq 0, \whs{0.5} b_i \geq a_i > 0, \whs{0.5} b_i > -a_i \geq 0, \whs{0.5} 
-a_i \geq b_i > 0,
\]
\begin{equation}
\label{octo}
-a_i > -b_i \geq 0, \whs{0.5} -b_i \geq -a_i > 0, \whs{0.5} -b_i > a_i \geq 0
\whs{0.5} a_i \geq -b_i > 0.
\end{equation}
This normalization of the normals makes it trivial to order them.
To decide whether the constraint's normal $N$ comes 
before $n$, or is equal to it,
we first compare the octants in which they lie. If they are in different
octants then we can order $N$ and $n$ by comparing the octants. If they
are in the same octant, it suffices to compare their ``secondary'' coordinate.
For instance, in the case when $a_i > b_i \geq 0$ we can simply compare
$n_y$ and $N_y$. As a result, the binary searches for finding
$A$ and $F$ require only the comparison of single numbers, and the
constant hidden in their $\wfc{O}{n \log n}$ complexity is small.
Moreover, by normalizing normals we reduce the cost of evaluating
the two products and a sum in $a_i x + b_i y$, which
compiles to a product and one fused add multiply (fma) instructions,
to the evaluation of $x + b_i y$, which compiles to
a single fma instruction. As a net effect, 
we exchange the cost of $n \wfc{O}{\log n}$ multiplications
by the cost of the  $2n$ divisions for normalization,
and this is a good deal for $n$ about one hundred.

Finally, we must be prepared to handle overflow
in the division $\wlr{-c_i} \oslash a_i$. Since we are rounding upwards,
we can only have $\wlr{-c_i} \oslash a_i = +\infty$. This implies that
$x + y$ must be larger than the largest finite floating point value, and
this contradicts our assumption \wref{bound} and leads us to the
conclusion that $\ws{F} = \emptyset$. Therefore, the algorithm terminates
when $\wlr{-c_i} \oslash a_i$ results in overflow.

%%%%%%%%%%%%%%%%%%%%%%%%%%%%%%%%%%%%%%%%%%%%%%%%%%%%
%% locating intersections
%%%%%%%%%%%%%%%%%%%%%%%%%%%%%%%%%%%%%%%%%%%%%%%%%%%%

\item[(ii)] In Figure \ref{figAlgo}, when searching for the location
of $X$ among the vertices $V  = \wlr{x,y} \in \wset{A,B,C,D,E,F}$,
for a new constraint as in Equation \wref{ci} we must evaluate the sign of
$a_i x + b_i y - c_i$, and for that we must represent
$x$ and $y$ somehow. If $V$ is the intersection of the consecutive edges
$a_j x + b_j y = c_j$ and $a_k x + b_k y = c_k$ 
in the counter clockwise order then 
\begin{equation}
\label{vertex_xy}
x = r/d \whs{1} \wrm{and} \whs{1} y = s/d
\end{equation}
for
\begin{equation}
\label{vertex}
r := c_j b_k - c_k b_j, \whs{0.7} 
s := a_j c_k - a_k c_j  \whs{0.7} \wrm{and} \whs{0.7}
d := a_j b_k - a_k b_j,
\end{equation}
and $x$ and $y$ cannot always be computed exactly with floating point arithmetic. 
Due to the ordering of the edges, $d > 0$ and we can bound
 $x$ and $y$ using
\begin{eqnarray}
\label{rsd1}
\wu{r} & := & c_k \otimes b_j \ \oplus \ \wlr{ \wlr{-c_j} \otimes b_k}, \\
\wu{s} & := & a_k \otimes c_j \ \oplus \ \wlr{ \wlr{-a_j} \otimes c_k}, \\
\wu{d} & := & a_k \otimes b_j \ \oplus \ \wlr{ \wlr{-a_j} \otimes b_k}, \\
\wo{r} & := & c_j \otimes b_k \ \oplus \ \wlr{ \wlr{-c_k} \otimes b_j}, \\
\wo{s} & := & a_j \otimes c_k \ \oplus \ \wlr{ \wlr{-a_k} \otimes c_j}, \\
\label{rsd6}
\wo{d} & = & a_j \otimes b_k \ \oplus \  \wlr{ \wlr{-a_k} \otimes b_j}. 
\end{eqnarray}
In these equations $u \oplus v$ is the value of $u + v$ rounded up
and $u \otimes v$ is $u * v$ rounded up. With this arithmetic,
we can prove that
\[
-\wu{r} \leq r \leq \wo{r}, \whs{1} -\wu{s} \leq s \leq \wo{s}
\whs{1} \wrm{and} \whs{1} -\wu{d} \leq d \leq \wo{d}.
\]
When $a_i \geq 0$, $b_i \geq 0$ and $c_i \geq 0$, we can analyze the sign of
 $a_i x + b_i y - c_i$ by comparing
\[
\wu{p} := - \wlr{ a_i \otimes \wu{r} \ \oplus \ b_i \otimes \wu{s}} 
\whs{1} \wrm{with} \whs{1}
\wo{q} := c_i \otimes \wo{d}.
\]
and
\[
\wo{p} := a_i \otimes \wo{r} \ \oplus \ b_i \otimes \wo{s} 
\whs{1} \wrm{with} \whs{1}
\wu{q} := -c_i \otimes \wu{d},
\]
and the analysis of the other of combinations of the signs of $a_i$, $b_i$ 
and $c_i$ is analogous.
If $\wu{p} > \wo{q}$, then certainly $a_i x + b_i y > c_i$,
and if $\wo{p} < \wu{q}$ then certainly $a_i x + b_i y < c_i$.
However if neither  $\wu{p} > \wo{q}$ nor $\wo{p} < \wu{q}$ then
we cannot obtain the sign of $a_i x + b_i y - c_i$ only 
with the information provided by the numbers \wref{rsd1}--\wref{rsd6}.
That is, the numbers \wref{rsd1}--\wref{rsd6} yield partial tests, which may be
inconclusive in rare cases. In these rare cases, we resort to
the technique presented in \cite{ES}. Using this technique
we can evaluate exactly the sign of
\begin{equation}
\label{exs}
a_i r + b_i s - c_i d = 
a_i c_j b_k - a_i  c_k b_j + 
b_i a_j c_k - b_i  a_k c_j + 
c_i a_k b_j  - c_i a_j b_k, 
\end{equation}
and decide on which side of the line $a_i x + b_i = c_i$ the vertex
$V$ lies. In summary, in order to locate $X$ we first
try our best with the numbers in Equations
\wref{rsd1}--\wref{rsd6}. If we fail then we resort
to the exact expressions for the vertex coordinates given by
Equations \wref{vertex_xy} and \wref{vertex} and  use
the more expensive evaluation of the sign of the expression
in Equation \wref{exs} with the technique described in \cite{ES}.

%%%%%%%%%%%%%%%%%%%%%%%%%%%%%%%%%%%%%%%%%%%%%%%%%%%%
%% computing intersections
%%%%%%%%%%%%%%%%%%%%%%%%%%%%%%%%%%%%%%%%%%%%%%%%%%%%

\item[(ii)] After finding in which edges $X$ and $Y$ lie, we must 
represent them somehow. Unfortunately, it is impossible
to always compute $X$ and $Y$ exactly with floating point
arithmetic. What we can do is to compute the numbers
in \wref{rsd1}--\wref{rsd6}, and use the equations 
\wref{vertex_xy} and \wref{vertex}  defining the
vertices as intersections of consecutive edges when the information
provided by these numbers are not enough.
\end{itemize}

\section{Representing the feasible region $\ws{F}$ in \texttt{C++}}
\label{sec_repr}

The building block we use to turn the ideas in the previous Section in C++ code
is the following struct, which we use to represent the edges illustrated
in Figure \ref{figRepr},
\begin{verbatim}
template <class T>
struct edge {
  T n;
  T c;
  T x[6];
};
\end{verbatim}

\begin{center}
\begin{figure}[!h]
\begin{tikzpicture}[scale = 1]

\filldraw[black] (0,0) circle(0.01pt);
\filldraw[black] (12,0) circle(0.01pt);
\filldraw[black] (0,4) circle(0.01pt);
\filldraw[black] (12,4) circle(0.01pt);

%        B     C            D         E          F        G         H      I           J         A
%\draw (2,1)--(4.64,1.44)--(5,2)--(4.9,2.6)--(4.6,3.4)--(3.6,4.3)--(2,4)--(1.3,3.2)--(1,2.5)--(1.4,1.6)--(2,1);

\draw (2.00 - 0.3 * 2.64, 1.000 - 0.3 * 0.44)--(4.64 + 0.3 * 2.64, 1.44 + 0.3 * 0.44); //  2.64,  0.44
\draw (4.64 - 0.6 * 0.36, 1.442 - 0.6 * 0.56)--(5.00 + 0.6 * 0.36, 2.00 + 0.6 * 0.56); //  0.36,  0.56
\draw (5.00 + 0.6 * 0.10, 2.000 - 0.6 * 0.60)--(4.90 - 0.6 * 0.10, 2.60 + 0.6 * 0.60); // -0.10,  0.60
\draw (4.90 + 0.5 * 0.30, 2.600 - 0.5 * 0.80)--(4.60 - 0.5 * 0.30, 3.40 + 0.5 * 0.80); // -0.30,  0.80
\draw (4.60 + 0.6 * 1.00, 3.400 - 0.6 * 0.90)--(3.60 - 0.6 * 1.00, 4.30 + 0.6 * 0.90); // -1.00,  0.90
\draw (3.60 + 0.5 * 1.60, 4.300 + 0.5 * 0.30)--(2.00 - 0.5 * 1.60, 4.00 - 0.5 * 0.30); // -1.60, -0.30  
\draw (2.00 + 0.5 * 0.70, 4.000 + 0.5 * 0.80)--(1.30 - 0.5 * 0.70, 3.20 - 0.5 * 0.80); // -0.70, -0.80
\draw (1.30 + 0.6 * 0.30, 3.200 + 0.6 * 0.70)--(1.00 - 0.6 * 0.30, 2.50 - 0.6 * 0.70); // -0.30, -0.70
\draw (1.00 - 0.6 * 0.40, 2.500 + 0.6 * 0.90)--(1.40 + 0.6 * 0.40, 1.60 - 0.6 * 0.90); //  0.40, -0.90
\draw (1.40 - 0.6 * 0.60, 1.600 + 0.6 * 0.60)--(2.00 + 0.6 * 0.60, 1.00 - 0.6 * 0.60); //  0.60, -0.60

\draw (2    + 0.1, 1    - 0.1)--(2    + 0.1, 1    + 0.1)--(2    - 0.1, 1    + 0.1)--(2    - 0.1, 1    - 0.1)--(2    + 0.1, 1    - 0.1);
\draw (4.64 + 0.1, 1.44 - 0.1)--(4.64 + 0.1, 1.44 + 0.1)--(4.64 - 0.1, 1.44 + 0.1)--(4.64 - 0.1, 1.44 - 0.1)--(4.64 + 0.1, 1.44 - 0.1);
\draw (5    + 0.1, 2    - 0.1)--(5    + 0.1, 2    + 0.1)--(5    - 0.1, 2    + 0.1)--(5    - 0.1, 2    - 0.1)--(5    + 0.1, 2    - 0.1);
\draw (4.9  + 0.1, 2.6  - 0.1)--(4.9  + 0.1, 2.6  + 0.1)--(4.9  - 0.1, 2.6  + 0.1)--(4.9  - 0.1, 2.6  - 0.1)--(4.9  + 0.1, 2.6  - 0.1);
\draw (4.6  + 0.1, 3.4  - 0.1)--(4.6  + 0.1, 3.4  + 0.1)--(4.6  - 0.1, 3.4  + 0.1)--(4.6  - 0.1, 3.4  - 0.1)--(4.6  + 0.1, 3.4  - 0.1);
\draw (3.6  + 0.1, 4.3  - 0.1)--(3.6  + 0.1, 4.3  + 0.1)--(3.6  - 0.1, 4.3  + 0.1)--(3.6  - 0.1, 4.3  - 0.1)--(3.6  + 0.1, 4.3  - 0.1);
\draw (2    + 0.1, 4    - 0.1)--(2    + 0.1, 4    + 0.1)--(2    - 0.1, 4    + 0.1)--(2    - 0.1, 4    - 0.1)--(2    + 0.1, 4    - 0.1);
\draw (1.3  + 0.1, 3.2  - 0.1)--(1.3  + 0.1, 3.2  + 0.1)--(1.3  - 0.1, 3.2  + 0.1)--(1.3  - 0.1, 3.2  - 0.1)--(1.3  + 0.1, 3.2  - 0.1);
\draw (1    + 0.1, 2.5  - 0.1)--(1    + 0.1, 2.5  + 0.1)--(1    - 0.1, 2.5  + 0.1)--(1    - 0.1, 2.5  - 0.1)--(1    + 0.1, 2.5  - 0.1);
\draw (1.4  + 0.1, 1.6  - 0.1)--(1.4  + 0.1, 1.6  + 0.1)--(1.4  - 0.1, 1.6  + 0.1)--(1.4  - 0.1, 1.6  - 0.1)--(1.4  + 0.1, 1.6  - 0.1);

\draw node at (2.00, 1.30){$B$};
\draw node at (4.64, 1.07){$C$};
\draw node at (5.30, 2.00){$D$};
\draw node at (4.60, 2.50){$E$};
\draw node at (4.30, 3.20){$F$};
\draw node at (3.60, 4.00){$G$};
\draw node at (1.80, 4.30){$H$};
\draw node at (1.60, 3.20){$I$};
\draw node at (1.30, 2.50){$J$};
\draw node at (1.15, 1.45){$A$};

\filldraw[black] (2.00, 1.00) circle(1.5pt); % B
\filldraw[black] (4.64, 1.44) circle(1.5pt); % C
\filldraw[black] (5.00, 2.00) circle(1.5pt); 
\filldraw[black] (4.90, 2.60) circle(1.5pt); % E
\filldraw[black] (4.60, 3.40) circle(1.5pt);
\filldraw[black] (3.60, 4.30) circle(1.5pt); % G
\filldraw[black] (2.00, 4.00) circle(1.5pt); % H
\filldraw[black] (1.30, 3.20) circle(1.5pt); % I
\filldraw[black] (1.00, 2.50) circle(1.5pt);
\filldraw[black] (1.40, 1.60) circle(1.5pt);

\draw node at (9.5,4.2){normalized normal vectors};
\draw node at (3.2,0.4){edges, vertices and boxes};

%(2,1)--(4.64,1.44) => (0.44, 2.64)
\draw node at (9.3,3.45){$BC$};
\draw[->] (9.5,2)--(9.5 - 0.4545 * 0.44,  2 + 0.4545 * 2.64);

%(3.6,4.3)--(2,4) => (0.3, -1.6)
\draw node at (9.8, 0.55){$GH$};
\draw[->] (9.5, 2)--(9.5 + 0.75 * 0.3, 2 - 0.75 * 1.6 );

%(1.3,3.2)--(1,2.5) => (0.7, -0.3)
\draw node at (11,1.45){$IJ$};
\draw[->] (9.5, 2)--(9.5 + 1.7142857 * 0.7, 2 - 1.7142857 * 0.3);

% (5,2)--(4.9,2.6) => (0.6, 0.1)
\draw node at (7.9, 1.8){$DE$};
\draw[->] (9.5,2)--(9.5 - 2 * 0.6, 2 - 2 * 0.1);

%(4.64,1.44)--(5,2) => (0.56, 0.36)
\draw node at (7.95,2.75){$CD$};
\draw[->] (9.5, 2)--(9.5 - 2.142857 * 0.56, 2 + 2.142857 * 0.36);

%(2,4)--(1.3,3.2) => (0.8, -0.7)
\draw node at (11 ,1){$HI$};
\draw[->] (9.5, 2)--(9.5 + 1.5 * 0.8, 2 -1.5 * 0.7);

%(1.4,1.6)--(2,1)=> (0.6, 0.4)
\draw node at (11, 2.75){$JA$};
\draw[->] (9.5,2)--(9.5 + 2 * 0.6, 2 + 2 * 0.4);

\draw node at (9.5 + 2.0 * 0.4 + 0.1, 2 + 2.0 * 0.6+ 0.2){$AB$};
\draw[->] (9.5,2)--(9.5 + 2.0 * 0.4, 2 + 2.0 * 0.6);

% (4.9,2.6)--(4.6,3.4) => (0.8, 0.3)
\draw node at (7.95, 1.4){$EF$};
\draw[->] (9.5,2)--(9.5 - 1.5 * 0.8, 2 - 1.5 * 0.3);

%(4.6,3.4)--(3.6,4.3) => (-0.9, =1)
\draw node at (9.5 - 2.0 * 0.4 - 0.2, 2 - 2.0 * 0.6 - 0.2){$FG$};
\draw[->] (9.5,2)--(9.5 - 1.2 * 0.9, 0.8);

\draw[dashed] (9.5 + 1.2, 2 - 1.2)--(9.5 + 1.2, 2 + 1.2)--(9.5 - 1.2, 2 + 1.2)--(9.5 - 1.2, 2 - 1.2)--(9.5 + 1.2, 2 - 1.2);

\draw[dashed] (9.5, 2)--(9.5 + 1.2, 2);
\draw[dashed] (9.5, 2)--(9.5 + 1.2, 2 + 1.2);
\draw[dashed] (9.5, 2)--(9.5 - 1.2, 2 - 1.2);
\draw[dashed] (9.5, 2)--(9.5 - 1.2, 2);
\draw[dashed] (9.5, 2)--(9.5,       2 - 1.2);
\draw[dashed] (9.5, 2)--(9.5,       2 + 1.2);
\draw[dashed] (9.5, 2)--(9.5 - 1.2, 2 + 1.2);
\draw[dashed] (9.5, 2)--(9.5 + 1.2, 2 - 1.2);
\end{tikzpicture}
\caption{$\ws{F}$ is stored in eight ranges of edges, which
correspond to the octants at the right.
The normals in each range are sorted in the counter clockwise order. 
An edge is given by the normalized  constraint and the six numbers
in Equations \wref{rsd1}--\wref{rsd6}. 
When $\wu{d}_i > 0$, these numbers yield a small box containing the $i$th vertex.
Even when $\wu{d}_i \leq 0$, they lead to quick tests to decide in
which side of a line this vertex is.
}
\label{figRepr}
\end{figure}
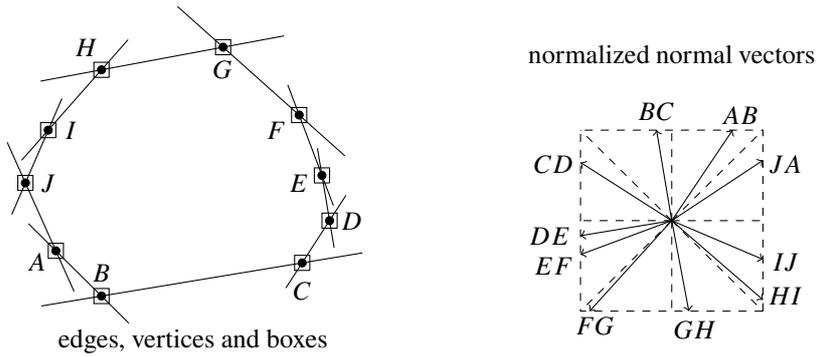
\end{center}

In the \texttt{struct} \texttt{edge<T>}, the field $\texttt{n}$ is the absolute value of the 
secondary entry of the normal
of the normalized constraint corresponding to the edge. 
Its interpretation depends on the
octant. In the first octant, the constraints are of the form $x + b_i y \geq c_i$,
and \texttt{n} is $b_i$. In the third octant, the constraints are $a_i x + y \geq c_i$,
and $\texttt{n}$ is equal to $-a_i \geq 0$.
In every octant, \texttt{c} is the right hand side of the 
constraint. The vector \texttt{x} contains the six numbers in Equations 
\wref{rsd1}--\wref{rsd6} corresponding to the first vertex in the edge. 
The symmetry among quadrants is perfect and we do not need to write 
an specific function to handle
edges in each octant, or to store the octant number in the edge.
Instead, we manipulate edges using functions of the form
\begin{verbatim}
template <int Octant, class T>
void function(edge<T> const& e)
\end{verbatim}
or 
\begin{verbatim}
template <int OctantA, int OctantB, class T>
void function(edge<T> const& ea, edge<T> const& eb)
\end{verbatim}
Due to symmetry, we can reason about such functions as if \texttt{Octant} = 0, 
or \texttt{OctantA} = 0, 
and let the compiler generate the code for all cases. In particular,
there is little need for switches to decide with which octant we are
working with at runtime. Most switches are performed at compile time.

Once we have decided to represent the edges and vertices 
of the feasible region by
the \texttt{struct} above we must choose how to store them in memory.
This choice depends upon how we expect to use the code. If the expected number
of edges $n_e$ of $\ws{F}$ is very large then it is advisable to use
a container in which insertion and removal of edges has cost
$\wfc{O}{\wfc{\log}{n_e}}$. However, these containers usually
have an overhead and are inefficient for small or even moderate
values of $n_e$.
For instance, the \texttt{C++} standard library provides a container
called \texttt{map} which is usually implemented as a red black tree
and is notoriously inefficient even for $n_e$ in the order
of a few hundred. There is also the subtle point that $n_e$ can be
much smaller than $n$, the number of inequalities. For instance,
if the constraints are generated randomly then our experiments
indicate that $n_e$ is much smaller than $n$ (something
like $\wfc{O}{\wfc{\log}{n}}$ seems plausible.)
For these reasons we organized the code in such way that 
we could replace the type of container with easy, and due to
time constraints we implemented only the container that 
we describe next. Insertion and removal of edges in this
container can cost $\wfc{O}{n_e}$ in the worst case, 
but it relies very little in dynamic memory allocation
and causes no fragmentation in memory. As a result, it
is quite efficient for the cases with $n_e$ up to a
hundred which concern us most.

The edges are grouped into eight \texttt{range<T>}s:

\begin{verbatim}
template <class T>
struct range {
  edge<T>* begin;
  edge<T>* end;
  range<T>* previous;
  range<T>* next;
};
\end{verbatim}

\newpage

\begin{center}
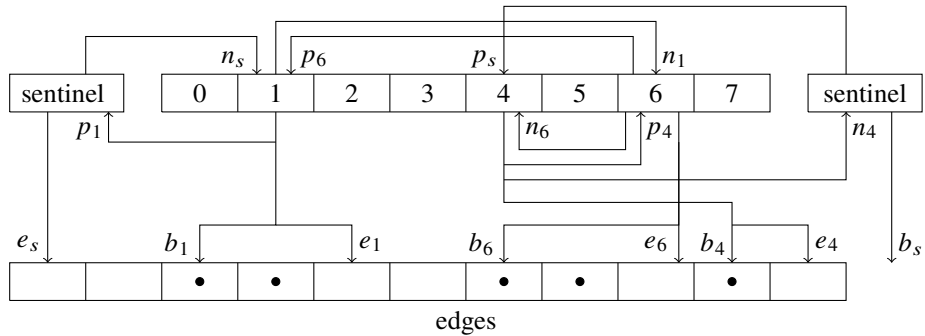
\begin{figure}[!h]
\begin{tikzpicture}[scale = 1]

\draw ( 0.0, 3.0)--(1.5, 3.0)--(1.5, 3.5)--(0.0, 3.5)--(0.0, 3.0);

\draw node at (0.7, 3.25){sentinel};
\draw[->](0.5, 3.0)--(0.5,1.0);
\draw[->] (1.0,3.5)--(1.0,4.0)-- (3.25, 4.0)--(3.25,3.5);
\draw node at (0.25, 1.25){$e_s$};
\draw node at (2.95,  3.7){$n_s$};

\draw ( 10.5, 3.0)--(12, 3.0)--(12, 3.5)--(10.5, 3.5)--(10.5, 3.0);
\draw node at (11.25, 3.25){sentinel};
\draw[->](11.6, 3.0)--(11.6,1.0);
\draw[->] (11.0,3.5)--(11.0,4.4)-- (6.5, 4.4)--(6.5,3.5);
\draw node at (11.85, 1.25){$b_s$};
\draw node at (6.25,   3.7){$p_s$};

\draw node at (2.5, 3.25){0};

\draw node at (3.5, 3.25){1};
\draw[->](3.5, 3.0)--(3.5,1.5)--(2.5,1.5)--(2.5,1.0);
\draw[->](3.5,1.5)--(4.5,1.5)--(4.5,1.0);
\draw[->](3.5, 2.6)--(1.3,2.6)--(1.3,3);
\draw[->](3.5, 3.5)--(3.5,4.2)--(8.5,4.2)--(8.5,3.5);

\draw node at (4.75, 1.25){$e_1$};
\draw node at (2.2, 1.25){$b_1$};
\draw node at (1.05, 2.75){$p_1$};
\draw node at (8.75, 3.7){$n_1$};

\draw node at (4.5, 3.25){2};
\draw node at (5.5, 3.25){3};

\draw node at (6.5, 3.25){4};
\draw[->] (6.5,2.3)--(8.3, 2.3)--(8.3,3);
\draw[->] (6.5, 2.1)--(11.0,2.1)--(11.0,3);
\draw[->] (6.5, 3.0)--(6.5, 1.8)--(9.5, 1.8)--(9.5,1.0);
\draw[->] (9.5,1.5)--(10.5,1.5)--(10.5,1.0);
\draw node at (9.25, 1.25){$b_4$};
\draw node at (10.75, 1.25){$e_4$};
\draw node at (11.25, 2.75){$n_4$};
\draw node at (8.55, 2.75){$p_4$};

\draw node at (7.5, 3.25){5};

\draw node at (8.5, 3.25){6};
\draw[->] (8.8, 3.0)--(8.8,1);
\draw[->] (8.2, 3.5)--(8.2, 4.0)--(3.7, 4.0)--(3.7, 3.5);
\draw[->] (8.1, 3.0)--(8.1,2.5)--(6.7,2.5)--(6.7,3);
\draw[->] (8.8, 2.6)--(8.8,1.5)--(6.5,1.5)--(6.5,1.0);
\draw node at (8.5, 1.25){$e_6$};
\draw node at (6.2, 1.25){$b_6$};
\draw node at (6.95, 2.75){$n_6$};
\draw node at (4.0,  3.7){$p_6$};

\draw node at (9.5, 3.25){7};

\draw ( 2.0, 3.0)--(10, 3.0)--(10, 3.5)--(2.0, 3.5)--(2.0, 3.0);
\draw ( 3.0, 3.0)--( 3.0, 3.5);
\draw ( 4.0, 3.0)--( 4.0, 3.5);
\draw ( 5.0, 3.0)--( 5.0, 3.5);
\draw ( 6.0, 3.0)--( 6.0, 3.5);
\draw ( 7.0, 3.0)--( 7.0, 3.5);
\draw ( 8.0, 3.0)--( 8.0, 3.5);
\draw ( 9.0, 3.0)--( 9.0, 3.5);

\draw ( 0.0, 0.5)--(11, 0.5)--(11, 1.0)--(0.0, 1.0)--(0.0, 0.5);
\draw ( 1.0, 0.5)--( 1.0, 1.0);
\draw ( 2.0, 0.5)--( 2.0, 1.0);
\draw ( 3.0, 0.5)--( 3.0, 1.0);
\draw ( 4.0, 0.5)--( 4.0, 1.0);
\draw ( 5.0, 0.5)--( 5.0, 1.0);
\draw ( 6.0, 0.5)--( 6.0, 1.0);
\draw ( 7.0, 0.5)--( 7.0, 1.0);
\draw ( 8.0, 0.5)--( 8.0, 1.0);
\draw ( 9.0, 0.5)--( 9.0, 1.0);
\draw (10.0, 0.5)--(10.0, 1.0);

\filldraw[black] (2.5,0.75) circle(1.5pt);
\filldraw[black] (3.5,0.75) circle(1.5pt);
\filldraw[black] (6.5,0.75) circle(1.5pt);
\filldraw[black] (7.5,0.75) circle(1.5pt);
\filldraw[black] (9.5,0.75) circle(1.5pt);

\draw node at (6, 0.2){edges};

\end{tikzpicture}
\caption{A feasible region with 5 edges, in ranges 1, 6, and 4.
The letters $b,e,p$ and $n$ stand for begin, end, previous and next. The sentinel
is represented twice but it is unique, and yes, it's begin comes after
it's end. The ranges 0, 2, 3, 5 and 7 are empty and
b = e = n = p = NULL for them.}
\label{figData}
\end{figure}
\end{center}

The ranges are managed by an object of type \texttt{ranges<T>}:

\begin{verbatim}
template <class T>
struct ranges{
	range<T> r[8];
	range<T> sentinel;
};
\end{verbatim}

Memory is organized as in Figure \ref{figData}.
Ranges can be inactive (when they are empty) or active.
The active ranges and the sentinel form a doubly linked
circular list, defined the field \texttt{previous} and \texttt{next}
in the ranges. They share a common array of edges, with the edges
for each active range being indicate by its \texttt{begin}
and \texttt{end} fields. For consistency with circularity, the \texttt{begin} field
of the sentinel always points to the first edge in the array of edges,
and the sentinel's \texttt{end} field always points to one passed
the last element of the array of edges.

We leave slack on the edges array so that removing an edge from a range
does not affect the other ranges and inserting an edge
in a range only affects it's neighboring ranges in a few cases. 
We can remove and insert edges in this data structure with
the usual techniques for the manipulation of arrays and
linked lists. When inserting an edge, if its range
is inactive then we search for largest gap between
the edges used by active ranges. If no gap is found
then we allocate an new array of edges twice as large as the current
one and move the current edges to it, dividing the
extra space roughly equally in between the space
used by the active ranges, and we activate the
range by inserting it into the list of active ranges.
If the new edge's range is active
and there is slack before or after it we simply
expand the set of edges of this range accordingly. 
If there is no slack then we try to bump one of its neighbors.
If this is not possible, then we reallocate memory
as before, and after that we expand the range.
To remove an edge we simply shrink the set of
edges managed by its range. If the range
becomes empty then we remove if from the
list of active ranges, but keep it as inactive 
in the \texttt{ranges<T>}'s array \texttt{r}.

\section{Testing}
\label{sec_test}

This section explains how to  generate test cases
for code that implements algorithms like the one we propose here.
We warn readers that it is quite hard to write correct code
for the task which we discuss in this article, and good
tests are essential to ensure the quality of our code. 
Tests in which we simply generate
constraints at random and verify that the resulting feasible region
is consistent are not enough. Such tests tend to generate feasible
regions with few edges, and will not find bugs caused by degenerate
cases like the ones in Figure \ref{figBug}. 

\begin{center}
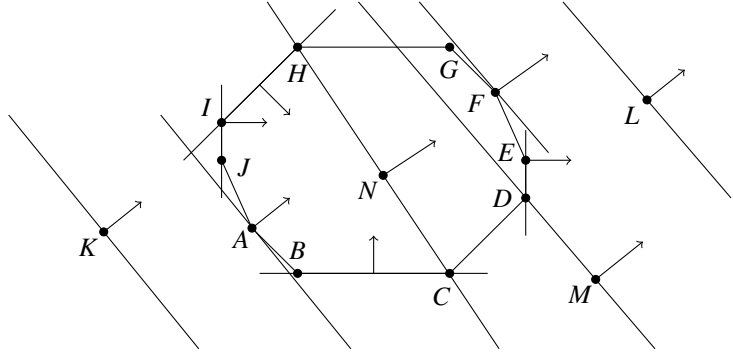
\begin{figure}[!h]
\begin{tikzpicture}[scale = 1]

\filldraw[black] (0,0) circle(0.01pt);
\filldraw[black] (12,0) circle(0.01pt);
\filldraw[black] (0,4) circle(0.01pt);
\filldraw[black] (12,4) circle(0.01pt);

\draw (6,1)--(8,1)--(9,2)--(9,2.5)--(8.6,3.4)--(8,4)--(6,4)--(5,3)--(5,2.5)--(5.4,1.6)--(6,1);

\draw node at (6,1.   3){$B$};
\draw node at (7.9,0.   7){$C$};
\draw node at (8.7,   2.0){$D$};
\draw node at (8.75, 2.6){$E$};
\draw node at (8.35, 3.25){$F$};
\draw node at (8,   3.7){$G$};
\draw node at (6,   3.65){$H$};
\draw node at (4.8,  3.2){$I$};
\draw node at (5.3, 2.4){$J$};
\draw node at (5.25,1.45){$A$};

\filldraw[black] (6,     1) circle(1.5pt);
\filldraw[black] (8,     1) circle(1.5pt);
\filldraw[black] (9,     2) circle(1.5pt);
\filldraw[black] (9,   2.5) circle(1.5pt);
\filldraw[black] (8.6, 3.4) circle(1.5pt);
\filldraw[black] (8,     4) circle(1.5pt);
\filldraw[black] (6,     4) circle(1.5pt);
\filldraw[black] (5,     3) circle(1.5pt);
\filldraw[black] (5,   2.5) circle(1.5pt);
\filldraw[black] (5.4, 1.6) circle(1.5pt);

% flat_max
\draw     (9, 2.9)--(9, 1.5);
\draw[->] (9, 2.5)--(9.6, 2.5);

% flat min
\draw     (5, 2)--(5, 3.5);
\draw[->] (5, 3)--(5.6, 3);

% HI
\draw (6.5, 4.5)--(4.5, 2.5);
\draw[->] (5.5, 3.5)--(5.5 + 0.4, 3.5 - 0.4);

% two points, N
\draw (5.6, 4.6)--(8.65,0.0);  
\filldraw[black] (7.125, 2.3)circle(1.5pt);
\draw[->] (7.125,2.3)--(7.125 + 0.15 * 4.6, 2.3 + 0.15 * 3.05);
\draw node at (7.125 -0.2,2.3 -0.2){$N$};

%  BC
\draw (5.5,1)--(8.5,1);
\draw[->] (7,1)--(7,1.5);

%M
\draw (6.8, 4.6)--(10.7,0.0);  
\filldraw[black] (10.7 - 3.9/5, 4.6/5) circle(1.5pt);
\draw node at (10.7 - 3.9/5 - 0.2, 4.6/5 - 0.2){$M$};
\draw[->] (10.7 - 3.9/5, 4.6/5)--(10.7 - 3.9/5 + 0.2 * 3.1, 4.6/5 + 0.2 * 2.5);

%L
\draw (11.7 - 1.3 * 1.7, 2 +  1.3 * 2)--(11.7,2.0);  
\filldraw[black] (11.7 - 0.65 * 1.7, 2 + 0.65 * 2 ) circle(1.5pt);
\draw node at (11.7 - 0.65 * 1.7 - 0.2, 2 + 0.65 * 2 - 0.2){$L$};
\draw [->] (11.7 - 0.65 * 1.7, 2 + 0.65 * 2)--(11.7 - 0.65 * 1.7 + 0.16 * 3.1, 2 + 0.65 * 2 + 0.16 * 2.5);

%A
\draw (4.2, 3.1)--(6.7,0);
\draw[->]  (5.4,1.6)--(5.4 + 0.16 * 3.1, 1.6 + 0.16 * 2.5);

% K
\draw [->] (3.45,1.55)--(3.45 + 0.16 * 3.1, 1.55 + 0.16 * 2.5);
\draw (2.2, 3.1)--(4.7,0); %% d = (2.5,-3.1) => 3.1, 2.5
\filldraw[black] (3.45,1.55) circle(1.5pt);
\draw node at (3.25,1.35){$K$};

\draw (7.6, 4.6)--(9.3,  2.6);
\draw[->] (8.6, 3.4)--(8.6 + 0.16 * 3.1 + 0.2, 3.4 + 0.16 * 2.5 + 0.1);

\end{tikzpicture}
\caption{Cases which lead to bugs in code implementing the algorithm described here.
The new constraints have a normal vector attached to them and
the current feasible region is the polygon $ABCDEFGHIJ$.
}
\label{figBug}
\end{figure}
\end{center}

In order to generate random test cases with an acceptable coverage 
we must direct the random choices. A reasonable family of test
cases can be built by choosing normals from the set of 32 equally space
normals on the border of the square $[-8,8] \times [-8,8]$.
\begin{equation}
\label{ns}
\ws{N} := 
\wset{ \wv{R}^{i}
\left(
\begin{array}{c}
8 \\
2 k
\end{array}
 \right)
\ \wrm{for} \ i = 0,1,2,3 \ \wrm{and} \ k = -4,-3,-2,\dots, 3},
\end{equation}
where 
\[
\wv{R} := 
\left(
\begin{array}{cc}
0 & -1 \\
1 & 0 
\end{array}
\right)
\]
is the counter clockwise rotation by $\pi/2$. With a powerful 
machine and multi threading we can test all ``valid'' subsets 
$\ws{S}$ of $\ws{N}$ as set of normals in the system \wref{lp}, 
where by valid we mean that there is no gap 
greater than or equal to $\pi$ between two consecutive
elements of $\ws{S}$ (if there is such a large gap then
the feasible region corresponding to $\ws{S}$ is unbounded.)

Given a valid subset $\ws{S} = \wset{\nu_0,\dots, \nu_{n-1}}$ 
of $\ws{N}$ with $n$ elements and an integer
$\beta  > 0$, in Subsection \ref{sec_sub} we explain how
to build a random polygon $\ws{P}$ with $n$ 
vertices $x_i,y_i \in \wds{Z} \cap [0,2^{\beta + 17})$
and numbers $\ell_i \in \wds{Z} \cap [1,2^{\beta + 12})$ such that
\begin{eqnarray}
x_{\wfc{s_n}{i}} & = & x_{i} + \ell_i \nu_{i,2}, \\
y_{\wfc{s_n}{i}} & = & y_{i} - \ell_i \nu_{i,1}, 
\end{eqnarray} 
where
\[
\wfc{s_n}{i} := \wlr{i+ 1} \wrm{mod} \ n.
\]
The $\ell_i$s are the lengths of the sides of $\ws{P}$ in the sup norm.
The polygon $\ws{P}$ is the feasible region of the problem \wref{lp} with
$a_i = \nu_{i,1}$, $b_i = \nu_{i,2}$. Since $\wnorm{\nu_i}_\infty = 2^3$, we have that
\begin{equation}
\label{aibici}
c_i = a_i x_i + b_i y_i \in \wds{Z} \cap [-2^{\beta + 21},2^{\beta + 21}].
\end{equation}
We can then execute the following procedure to test all cases
in Figure \ref{figBug} (except for the ones in which the
new constraint contains two no adjacent vertices of $\ws{P}$)
using a floating point arithmetic in which the mantissa has
$\beta + 22$ bits and are such that $2^{\beta + 30}$
does not overflow. For instance, for $\beta \leq 1$  we could
use \texttt{float}, and for $\beta \leq 30$ we could use
\texttt{double}. 

To start with, we pick randomly a few orders on the 
indexes $i \in \wset{0,\dots, n- 1}$ and for each order, starting from the square
$[0,2^{\beta + 30}] \times [0,2^{\beta + 30}]$, we insert the
constraints in Equation \wref{aibici} one by one, checking whether
the current feasible region is consistent at each step.
In the end we check whether the final feasible region is $\ws{P}$.
We then consider the set $\wo{\ws{N}}$ of 64 normals equally spaced on the border of the square
$[-8,8] \times [-8,8]$
\[
\wo{\ws{N}} := 
\wset{ \wv{R}^{i}
\left(
\begin{array}{c}
8 \\
k
\end{array}
 \right)
\ \wrm{for} \ i = 0,1,2,3 \ \wrm{and} \ k = -8,-7,-6,\dots, 7}.
\]
For each $\nu \in \wo{\ws{N}}$ we compute the $n$ values
\[
h_i := \nu_{i,1} x_i + \nu_{i,2} y_i \in [-2^{\beta + 21},2^{\beta + 21}]
\]
and let $\ws{U} = \wset{u_0, u_1, \dots u_{m-1}}$ be the set we obtain after we sort 
$\ws{H} := \wset{h_0,h_1,\dots,h_{n-1}}$ and remove the repetitions. 
For each $u_i \in \ws{U}$ we check whether the constraint
\[
\nu_{i,1} x + \nu_{i,2} y \geq u_i
\]
is inserted correctly. We then define $u_{-1} = -2^{\beta  + 22}$ 
and $u_{m} = 2^{\beta + 22}$ and for 
$i = -1, \dots, m$ we choose a few $u$ randomly in 
$\wlr{u_{i},u_{i+1}}$ and check whether the constraint
\[
\nu_{i,1} x + \nu_{i,2} y \geq u
\]
is inserted correctly. 

The tests above do no cover degenerate $\ws{F}$s, i.e.,
the cases in which $\ws{F}$ is empty, a point or a segment
(In our code, we represent such cases using another data structure.)
We can generate tests
for these cases by a procedure similar to the one above, but which is less
time consuming. To test the case in which $\ws{F}$ is a point,
we generate such point as the intersection
of two segments with the normals in the set $\ws{N}$ in 
Equation \wref{ns} and $c_i$ chosen randomly so that the resulting
point has non negative coordinates. To generate test cases for
segments, we generate them using 3 elements from $\ws{F}$
as normals and $c_i$'s chosen as for points. We then
check the insertion of the same new constraints as for the
case in which $\ws{P}$ is a polygon.

In our experience, the test procedures above
are quite powerful: using them we found bugs in our code which were not
found by our unit tests which focused on each small part 
of the code at a time, due to incorrect implicit assumptions
we made while coding these unit tests.

\subsection{Building a polygon given its normals}
\label{sec_sub}
This subsection describes how to generate 
a polygon $\ws{P}$ with edges with normals in a valid subset 
$\ws{S} = \wset{\nu_0, \nu_2, \dots \nu_{n-1}}$ of the set
of normals $\ws{N}$ in Equation \wref{ns}.
We assume $\ws{S}$ to be sorted in the counter clockwise order.
We now define the successor of $v_i \in \ws{S}$ as
\[
\sigma_i := \nu_{\wfc{s_n}{i}},
\]
and find the lengths $\ell_i$ of the edges of $\ws{P}$ in the sup norm.
We generate $n$ random integers $t_p \in [1,2^{\beta})$ and compute the sum
\[
\Delta := \sum_{p = 0}^{n - 1} t_p v_p.
\]
The lengths $t_p$ define the sides of a closed polygonal line if and
only if $\Delta = 0$. In the unlikely case that $\Delta = 0$, we happily set
$\ell_p = t_p$ for all $p$ and go for coffee.
Otherwise, we fix the $t_p$'s as follows.
Since there are at most $4$ elements in each octant in $\ws{N}$
and $\wnorm{\nu_i}_\infty = 8 = 2^{3}$ for all $i$, we can write
\[
\delta_2 := \sum_{i = 0}^3 \sum_{j = 0}^{3} z_{i,j,2}
- \sum_{i = 4}^7 \sum_{j = 0}^{3} z_{i,j,2}
\]
for integers $z_{i,j,q} \in [0,2^{3}]$, and
\begin{equation}
\label{bd}
\wabs{\delta_2} < 2^{\beta + 7}.
\end{equation}
By symmetry,
$\wabs{\delta_1} < 2^{\beta + 7}$.
Let $j \in [0,n)$ be such that $\nu_j \leq -\Delta < \sigma_j$ 
in the counter clockwise order in which $\ws{S}$ is sorted.
Since $\ws{S}$ is valid, there exist $\alpha_\nu,\alpha_\sigma \geq 0$ such that
\[
-\Delta = \alpha_\nu \nu_j + \alpha_\sigma \sigma_j.
\]
This implies that $\alpha_\nu = \tilde{\alpha}_\nu/d$ 
and $\alpha_\sigma = \tilde{\alpha}_\sigma/d$ 
with 
\begin{eqnarray}
\tilde{\alpha}_\nu     & := &  \delta_2 \, \sigma_{j,1} - \delta_1 \, \sigma_{j,2}, \\ 
\tilde{\alpha}_\sigma  
& := &  \nu_{j,2} \, \delta_1   - \nu_{j,1} \, \delta_2,  \\ 
d                      & := & \nu_{j,1} \, \sigma_{j,2} - \nu_{j,2} \, \sigma_{j,1}.
\end{eqnarray}
$d$ is positive due to the order in $\ws{S}$. Since
$\wnorm{\nu_j}_\infty = \wnorm{\sigma_j}_\infty = 2^3$,
$d \in \wds{Z} \cap [1, 2^7]$. Similarly, the bound \wref{bd} yields
\[
\tilde{\alpha}_\nu,\ \tilde{\alpha}_\sigma  
\in \wds{Z} \cap [0, 2^{\beta + 11}).
\]
It follows that 
\begin{eqnarray}
\label{lenp}
\ell_p & := & d t_p \whs{1.2} \in  \wds{Z} \cap
 [1, 2^{\beta + 7})
\whs{0.3} \wrm{for} \ p \not \in \wset{j,\wfc{s_n}{j}},\\
\label{lj}
\ell_j & := &  d t_j + \tilde{\alpha}_\nu 
\whs{0.5}   \in  \wds{Z} \cap
[1, 2^{\beta + 12}),
\\
\label{lsn}
\ell_{\wfc{s_n}{j}} & := & d t_{\wfc{s_n}{j}} + \tilde{\alpha}_\sigma 
 \whs{0.05}  \in \wds{Z} \cap
 [1, 2^{\beta + 12}),
\end{eqnarray} 
and $\sum_{p = 0}^{n - 1} \ell_p \nu_p = 0$.
We now have the normals and the lengths for the sides of $\ws{P}$ and
build the $x_i$ and $y_i$ in two steps.
First we define 
\[
q_p := \wlr{\wfc{s_n}{j} + p} \! \! \! \mod n
\]
and then, for $k = 0,1,\dots,n - 1$, 
\begin{equation}
\tilde{x}_{k} :=  \sum_{p = 1}^k \ell_{q_p} \nu_{q_p,2}
\whs{1} \wrm{and} \whs{1}
\tilde{y}_{k} := -\sum_{p = 1}^k \ell_{q_p} \nu_{q_p,1},
\end{equation}
with the usual convention that $\sum_{p=1}^0 u_p = 0$.
The identity $\sum_{p = 0}^{n-1} \ell_p \nu_p =0$ leads to
\[
\tilde{x}_{n-1} = - \ell_j \nu_{q_{n-1},2}
\whs{1} \wrm{and} \whs{1}
\tilde{y}_{n-1} = \ell_j \nu_{q_{n-1},1},
\]
and Equations \wref{lj} and \wref{lsn} and the fact that 
$\wnorm{\nu_{q_{n-1}}}_\infty = 2^3$ imply that 
\begin{equation}
\label{bx1}
\wabs{\tilde{x}_{n-1}} < 2^{\beta + 15}
\whs{1} \wrm{and} \whs{1}
\wabs{\tilde{y}_{n-1}}  <  2^{\beta + 15}.
\end{equation}
Applying the same argument used to obtain the bound
$\delta_2$ in Equation \wref{bd} with $\ell_p$ in Equation
\wref{lenp} instead of $t_p$ we obtain that
\begin{equation}
\label{bx2}
\wabs{\tilde{x}_k} <  2^{\beta + 14}
\whs{0.5} \wrm{and} \whs{0.5}
\wabs{\tilde{x}_k} <  2^{\beta + 14}
\whs{0.5} \wrm{for} \whs{0.5} k = 0,1,\dots,n - 2.
\end{equation}
Equations \wref{bx1} and \wref{bx2} imply that
\[
\wu{x} := \min_{i \leq 0 < n} \tilde{x}_i 
\whs{1} \wrm{and} \whs{1} 
\wu{y} := \min_{i \leq 0 < n} \tilde{y}_i 
\]
have absolute value smaller than $2^{\beta + 15}$.
We then flip our last coins to find integers
$\delta_x,\delta_y \in [0,2^{\beta + 16})$
and define
\begin{eqnarray}
\nonumber
x_k := \tilde{x}_{\wlr{n + k - \wfc{s_n}{j}} \! \! \! \mod n} - \wu{x} + \delta_x
\in \wds{Z} \cap [0,2^{\beta + 17}), \\
\nonumber
y_k := \tilde{x}_{\wlr{n + k - \wfc{s_n}{j}} \! \! \! \mod n} - \wu{y} + \delta_y
\in \wds{Z} \cap [0,2^{\beta + 17}),
\end{eqnarray}
completing the construction of $\ws{P}$.

\end{document}